%% file: conference_101719.tex
\documentclass[conference]{IEEEtran}
\IEEEoverridecommandlockouts
\usepackage{cite}
\usepackage{amsmath,amssymb,amsfonts}
\usepackage{algorithmic}
\usepackage{graphicx}
\usepackage{textcomp}
\usepackage{xcolor}
\usepackage{graphicx}
\usepackage{multicol}
\input{acro.tex}
\usepackage{float}

\def\BibTeX{{\rm B\kern-.05em{\sc i\kern-.025em b}\kern-.08em
    T\kern-.1667em\lower.7ex\hbox{E}\kern-.125emX}}
\begin{document}

\title{An Efficient and Flexible  Deep Learning Method for Signal Delineation via Keypoints Estimation}

\author{
\IEEEauthorblockN{1\textsuperscript{st} Adrian Atienza *}
\IEEEauthorblockA{\textit{Department of Health Technology} \\
\textit{Technical University of Denmark (DTU)}\\
Copenhagen, Denmark \\
adar@dtu.dk}

\and

\IEEEauthorblockN{2\textsuperscript{nd} Jakob Eyvind Bardram}
\IEEEauthorblockA{\textit{Department of Health Technology} \\
\textit{Technical University of Denmark (DTU)}\\
Copenhagen, Denmark \\
jakba@dtu.dk}

\and

\IEEEauthorblockN{3\textsuperscript{nd} Peter Karl Jacobsen}
\IEEEauthorblockA{\textit{Department of Cardiac Diseases} \\
\textit{Rigshospitalet}\\
Copenhagen, Denmark \\
peter.karl.jacobsen@regionh.dk}

\and

\IEEEauthorblockN{4\textsuperscript{nd} Sadasivan Puthusserypady}
\hspace{19.5cm}\IEEEauthorblockA{\textit{Department of Health Technology} \\
\textit{Technical University of Denmark (DTU)}\\
Copenhagen, Denmark \\
sapu@dtu.dk}
}
\maketitle

\begin{abstract}
\acf{DL} methods have been used for \acf{ECG} processing in a wide variety of tasks, demonstrating good performance compared with traditional signal processing algorithms.
These methods offer an efficient framework with a limited need for apriori data pre-processing and feature engineering.
While several studies use this approach for ECG signal delineation, a significant gap persists between the expected and the actual outcome.
Existing methods rely on a sample-to-sample classifier. However, the clinical expected outcome consists of a set of onset, offset, and peak for the different waves that compose each R-R interval.
To align the actual with the expected output, it is necessary to incorporate post-processing algorithms. This counteracts two of the main advantages of \ac{DL} models, since these algorithms are based on assumptions and slow down the method's performance.
In this paper, we present \ac{KEED}, a novel \ac{DL} model designed for keypoint estimation, which organically offers an output aligned with clinical expectations. 
By standing apart from the conventional sample-to-sample classifier,  we achieve two benefits: (i) Eliminate the need for additional post-processing, and (ii) Establish a flexible framework that allows the adjustment of the threshold value considering the sensitivity-specificity tradeoff regarding the particular clinical requirements.
The proposed method's performance is compared with \ac{SOTA} signal processing methods. Remarkably, \ac{KEED} significantly outperforms despite being optimized with an extremely limited annotated data. In addition, \ac{KEED} decreases the inference time by a factor ranging from 52x to 703x.
\end{abstract}
\begin{IEEEkeywords}
\Acf{ECG} delineation, P-wave identification, keypoints estimation model, \acf{DL}.
\end{IEEEkeywords}

\section{Introduction}

\begin{figure}[t]
\begin{center}
\centerline{\fbox{\includegraphics[width=0.7\columnwidth]{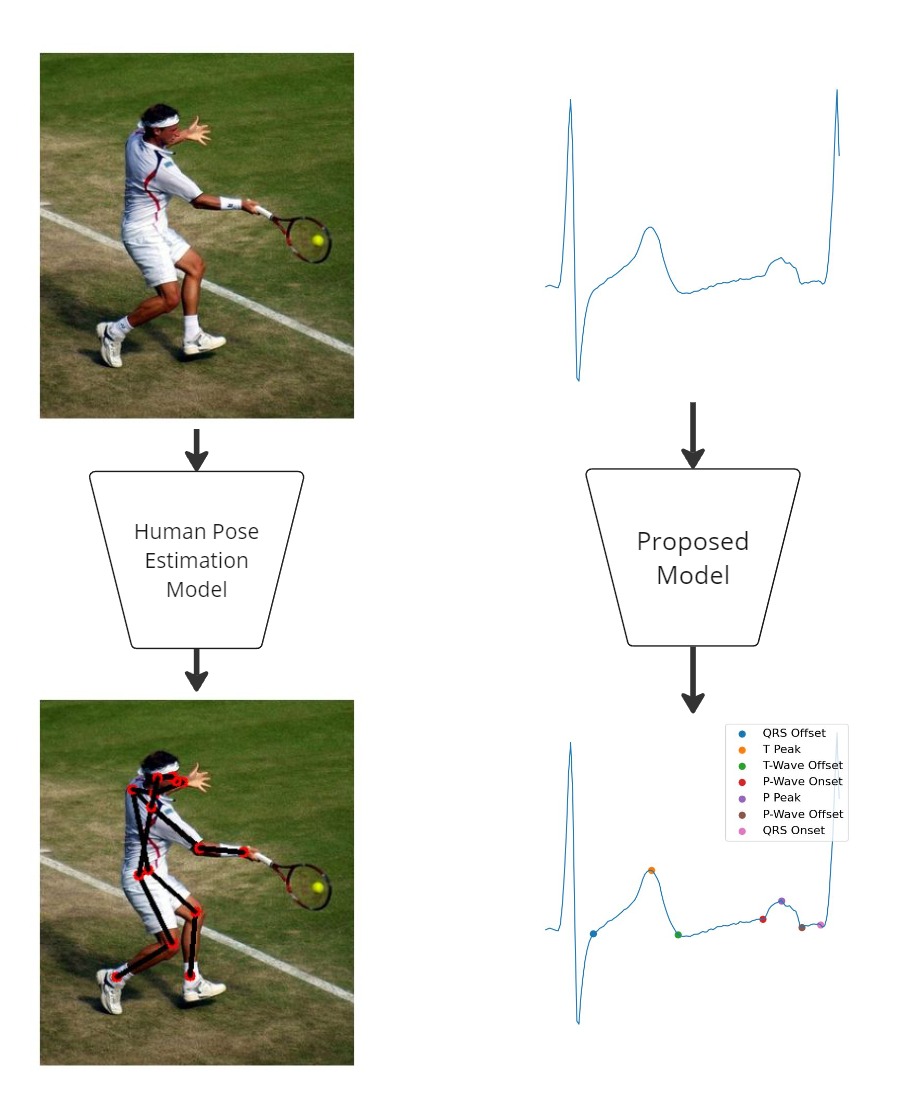}}}
\caption{Analogy between Human Pose Estimation Task and Signal Delineation task.}
\label{fig:shared_framework}
\end{center}
\end{figure}

When processing \ac{ECG} recordings, the presence, location, and morphology of different component waveforms (P, QRS, and T) within the R-R interval is of high clinical relevance due to several reasons:
(i) Discretising between ventricular and atrial beats is of major importance for an accurate beat rate calculation for arrhythmias identification, 
(ii) The absence of P-waves serves as a crucial indicator in identifying \acf{AFib}  \cite{afib1, afib2},
(iii) T-wave morphology plays a crucial role in the early identification and diagnosis of ventricular arrhythmias\cite{twave1, twave2}.

Considering the significance of this task, it is unsurprising that numerous studies have investigated its automation. These studies can be divided into either conventional signal processing or \ac{DL} methods.
The gold standard signal processing methodology \cite{wavelet} involves a delineation system based on \ac{WT}, where the set of peak, onset, and offset of each P-wave, T-wave, and QRS complex is determined by its waves morphology and peaks. 
Unlike traditional signal processing methods, \ac{DL} approaches eliminate the need for manual feature engineering. Existing methods \cite{model1, model2} classify each sample as P-wave, QRS complex, T-wave, and No wave. \\

\ac{DL} offers a framework with a lack of assumptions. It means that a need for algorithm reparameterization when it under-performs on specific outliers can be avoided by incorporating these outliers into the training set. Additionally, the inherent parallel computing efficiency of these models provide a significant advantage compared with traditional signal processing methods.
However, there is a gap between the current outcome of existing \ac{DL} methods and a clinician's expectation. While these methods return a sample-to-sample classification, the professionals expect to obtain the presence or absence of both the T-wave and P-wave for each R-R interval, with its respective location. 
Even though it is possible to move the DL model outcome closer to the expectation, this requires post-processing algorithms, leading to a decrease in the efficiency of these models. In addition, these algorithms require a hyperparameter configuration based on assumptions. \\

In this paper, we present \acf{KEED}, specifically designed for \ac{ECG} signal delineation.
We find the human pose estimation task and its analogy with signal delineation, as illustrated in Figure~\ref{fig:shared_framework}.
An R-R interval as well as the human skeleton, comprises distinct points that may be either present or absent. Moreover, the position of each of these points assists in locating subsequent ones. For instance, similar to how the right wrist is proximate to the right elbow within the human skeletal structure, the T-wave onset precedes the T-peak and the T-offset.
By adopting this approach, \ac{KEED} not only returns the expected outcome by clinicians without further processing but also provides a flexible framework that can be fine-tuned to strike the right balance between sensitivity and specificity based on the specific clinical context.
We have evaluated \ac{KEED} performance against the gold standard signal processing \ac{WT}-based methods in publicly available databases, which demonstrated significantly better performance than theirs. In addition, due to both the parallel computing inherent in \ac{DL} modeling and the lack of further post-processing, the inference time is reduced by 52x to 703x when executed on a local computer.\\

In summary, the contributions of this paper are: 
\begin{itemize}
\item Introduced \ac{KEED}, a novel \ac{DL} method that takes apart from existing sample-to-sample classifier by addressing the signal delineation task from a keypoint estimation approach. By taking this approach, the model’s output aligns with clinicians’ expected outcomes without requiring additional processing.

\item \ac{KEED} organically offers the possibility to set up a $\lambda$ parameter that can be adjusted to the Sensitivity vs Specificity trade-off required for a particular clinical setup.

\item The \ac{KEED} model outperforms existing signal processing methods, significantly reducing inference time. Additionally, given that this model was trained with a limited amount of labeled data, we can anticipate that its performance will improve as more labeled data is incorporated during optimization.

% \item We open a venue for adapting signal delineation modeling to a keypoint estimation task during model design and optimization.
\end{itemize}

\begin{figure*}[t]
\begin{center}
\centerline{\fbox{\includegraphics[width=\linewidth]{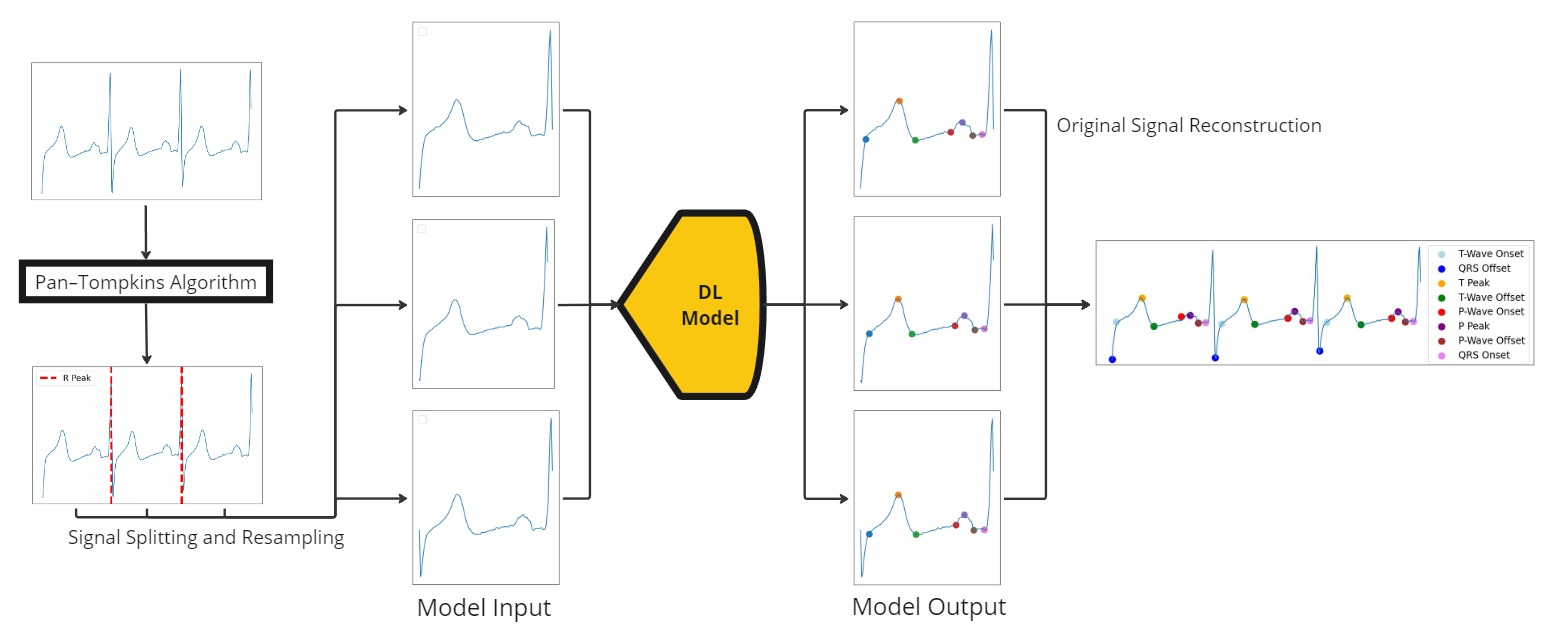}}}
\caption{\acf{KEED} is displayed. The \ac{ECG} recording is processed by the Pan–Tompkins algorithm for R peak identification. The signal is split into R-R intervals which are passed through the DL model. After discretizing the presence/abscense of each keypoint based of the computed probability and the $\lambda$ parameter, the locations of the present keypoints are translated to match the original input.}
\label{fig:keed}
\end{center}
\end{figure*}

\section{Related Work}
\subsection{Wavelet Methods for Signal Delineation}
The wavelet based method\cite{wavelet} is the gold standard signal processing algorithm for \ac{ECG} delineation task. This particular study introduces a robust single-lead ECG delineation system using the \acf{WT} decomposition,
where the signal is decomposed into a set of basis functions. 
This approach entails a manual feature engineering process for configuring distinct thresholds utilized by the algorithm. Additionally, it not only requires performing the \ac{WT} decomposition but also sequentially processing the entire signal.
The approach proposed in this study circumvents the need for manual feature engineering. Instead, it provides a framework that operates without any prior assumptions. Moreover, each R-R interval is processed in parallel, resulting in a significantly decreased inference time. The outcome computed by both methods is equivalent and will be evaluated in Section 4.

\subsection{\acf{DL} Methods for Signal Delineation}
Existing \ac{DL} methods such as DENS-ECG \cite{model1} propose a model that combines convolutional layers and long short-term memory (LSTM) layers to classify each sample as P-wave, QRS complex, T-wave, and No wave. 
This classification at the sample level does not correspond to the expected outcomes by clinicians, which involve identifying the onset, offset, and peak of each component waveforms.
While transitioning from sample-to-sample classification to keypoint clinical expected output is feasible, it necessitates a post-processing procedure that involves analyzing the entire sequence of classifications. These post-processing methods are susceptible to fragility due to potential sample misclassifications. To address this concern, alternative studies, such as the work by \cite{model2}, propose a more robust post-processing approach.
%
%These post-processing methods slow down inference time and are made up of different assumptions. In this paper, we present a \ac{DL} method that does not need these post-processing methods.
%
These post-processing techniques lead to increased inference time due to their reliance on various assumptions. In this paper, we introduce a \ac{DL} approach that completely eliminates the need for such post-processing methods, which makes it a much faster method.

\subsection{Human-Pose Estimation models}
A significant contribution of this paper lies in addressing the signal delineation challenge by focusing on the estimation of keypoints. A prominent example within this domain is the Human Pose Estimation task. Existing studies, i.e., \cite{hourglass, model}, typically approach this task in a two-step process. First, an auxiliary model is employed to identify individuals within the image. Subsequently, the cropped regions corresponding to each person are fed into a keypoint model for precise human pose estimation. 
The model architecture corresponds to a U-Net-based model. The output of this model consists of an expansion of input by K channels, where K represents the number of keypoints the model is trained to recognize. Each channel corresponds to the probability of each keypoint being present at a specific coordinate.

\section{\acf{KEED}}
The schematic of the proposed \ac{KEED} is Illustrated in Figure \ref{fig:keed}.
The Pan–Tompkins algorithm \cite{pantomkins} is used for finding R peaks of the whole \ac{ECG} input to
split it into R-R intervals.
All R-R intervals are resampled to a common length to match the input size of the model before
being passed through the \ac{DL} model. This model returns the location of each keypoint in addition to its associated probability. 
The $\lambda$ hyperparameter is used as a threshold to asses the absence/presence of each keypoint based on its probability.
Each of the R-R interval keypoints locations are reconstructed to match the original input.

\subsection{Analogy with Human-Pose Estimation Models}
KEED is closely tied to how the state-of-the-art (SOTA) methods tackle the challenge of human pose estimation. These methods initially identify individuals within an image using a primary model. Subsequently, the cropped regions corresponding to each person are fed into the main model responsible for locating skeleton keypoints. The analogy of this process can be visualized through Figure~\ref{fig:keed}. \ac{KEED} leverages the Pan–Tompkins algorithm to pinpoint R-R intervals, which are then processed in parallel by the \ac{DL} model. Once each R-R keypoints localization is computed, these coordinates are translated to match the original input.

\subsection{\acf{DL} Model Architecture}
The parallelism between \ac{KEED}’s approach for signal delineation and the way SOTA methods address the human pose estimation task leads us to adapt an existing model from the latter domain for \ac{ECG} signal delineation.
In particular, the selected model is an improved version of the HourGlass \cite{hourglass} architecture proposed in the study \cite{model}. It combines a certain number of blocks of the widely used U-Net architecture with the efficient, so-called  ``soft-gated skip connections'' to achieve a notable performance in the human pose estimation task.

\begin{figure*}[t]
\begin{center}
\centerline{\fbox{\includegraphics[width=0.65\linewidth]{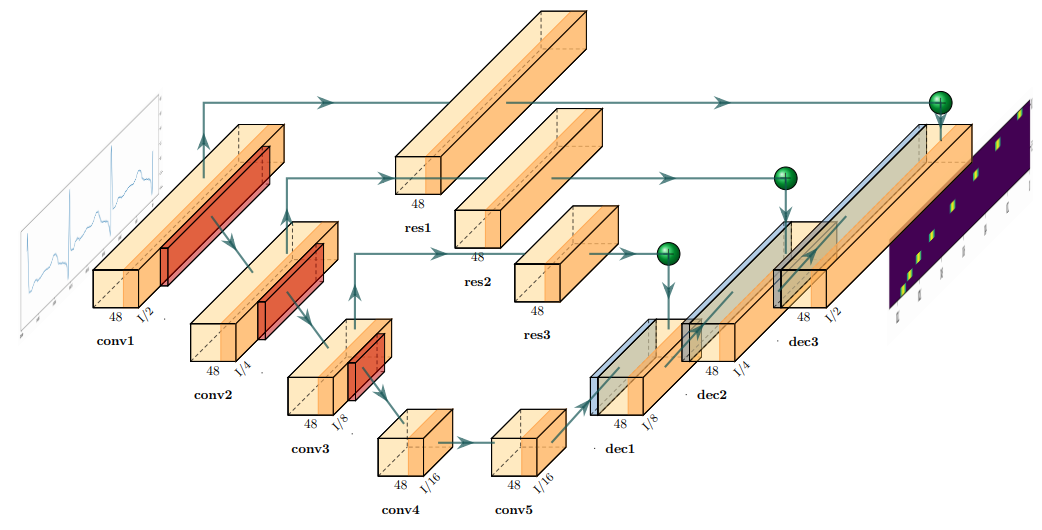}}}
\caption{The \ac{DL} U-Net-based architecture used is displayed. It consists on an encoder which synthesises the information contained in the input within a dense latent space. This space is used by a decoder for the reconstruction of an equivalent-dimension output expanded with K channels, being K the number of keypoints to be identified. Each output channel represents the probability of the respective keypoint to be located in each sample. Both encoder and decoder are linked through residual connections. }
\label{fig:model}
\end{center}
\end{figure*}

To adapt this model to ECG processing, the 2D convolutional layers have been replaced by 1D, fitting the ECG signal dimension. Figure~\ref{fig:model} illustrates the diagram of each block. It can be seen how the ECG strip is fed into the U-Net-based model to compute the probabilities of each keypoint to be located in each sample. The $\lambda$ parameter determines the threshold at which each keypoint transitions from being absent to being present, based on its associated probability. 

\subsection{Model Implementation and Training}
The proposed \ac{KEED} implementation consists of the following hyperparameter configuration: (i) The model dimension is set to 48 features for each layer, (ii) We use 4 layers in both the encoder and decoder components of the U-Net,  (iii) We use 2 U-Net blocks, 
 and (iv) the $\lambda$ parameter is set up to 0.4.

\ac{KEED} has been optimized using only the 2877 manually labeled beats belonging to the \acf{QT}. Note that 171 out of these 2877 overlap with the \acf{MIT-PWave} dataset which has been used for the evaluation. Therefore, these overlapping beats have been removed from the training set. We use a batch size of 64, and Adam \cite{adam} with a learning rate of $0.001$ and a weight decay of $1e-6$ as the optimizer.

\section{Evaluation}
For assessing the model performance, we have performed up to three evaluations, involving two distinct databases from the one that has been used for training the model. In addition, the impact of the $\lambda$ parameter has been evaluated, to prove that this parameter should be adjusted depending on the particular clinical specifications.

\subsection{\Acf{SOTA} Evaluation}
To assess KEED’s performance relative to existing methods, we conducted two distinct experiments using separate, non-overlapping datasets distinct from the one used during training. The used datasets consist of \acf{MIT-PWave}\cite{mit-p} and \ac{BUT PDB}\cite{brno}, which are publicly available in Physionet\cite{physionet}. The performance of the proposed method is compared with \ac{WT} method, as its outcome is comparable to that of KEED. The Neurokit implementation\cite{neurokit2} with its three distinct variations of the method (Continuous Wavelet Transform (CWT), Discrete Wavelet Transform (DWT) and Peak) are used in the evaluation. In both experiments, we calculate the corresponding metrics regarding the presence/absence of the P-Wave in each R-R interval. The calculated metrics consist on the standard accuracy, sensitivity and specificity ratios. In addition, we present an error metric that represents the distance in sample units between the predicted peak and the labeled peak (when both are present). Finally, we evaluate the inference time required for each method.\\

\textbf{\ac{MIT-PWave} evaluation:} This database contains reference P-wave annotations for twelve signals. Since the distinct recordings belong to subjects suffering sporadic \ac{AFib} episodes, it is expected that the P-wave will not be present in all R-R intervals. Table~\ref{tab:mit-afib-p} represents the obtained metrics, where it can be seen that the proposed method outperforms the distinct variations of \ac{WT} signal processing method. Remarkably, it can be seen how the inference time required for processing the whole database is of the order from 52x to 703x faster when computed in a local machine with a 2080 Nvidia RTX GPU card.
    
\begin{table}[H]
\caption{Evaluation on \ac{MIT-PWave} dataset. Best metrics are highlighted in bold type.}
\resizebox{\columnwidth}{!}{%
\label{tab:mit-afib-p}
\begin{tabular}{|c|ccccc|}
\hline
Method & \begin{tabular}[c]{@{}c@{}}Accuracy \\ (\%)\end{tabular} & \begin{tabular}[c]{@{}c@{}}Sensitivity\\  (\%)\end{tabular} & \begin{tabular}[c]{@{}c@{}}Specificity \\ (\%)\end{tabular} & \begin{tabular}[c]{@{}c@{}}Error\\ (Samples)\end{tabular} & \begin{tabular}[c]{@{}c@{}}Time \\ Consumed (s)\end{tabular} \\ \hline
Peak   & 89.9 & 98.0 & 92.3 & 5.7 & 192.6 \\
CWT    & 88.9 & 95.8 & 92.3 & 45.1 & 2039.6 \\
DWT    & 91.9 & 99.7 & 92.1 & 14.8 & 151.5 \\ \hline
KEED   & \textbf{96.4} & 98.0 & \textbf{98.0} & 7.5 & \textbf{2.9} \\ \hline
\end{tabular}}
\end{table}

\textbf{\ac{BUT PDB} evaluation:} This database consists of 50 2-minute 2-lead ECG signal records with various types of pathology. Table~\ref{tab:brno} represents the experiment results. 
    
\begin{table}[H]
\caption{Evaluation on \ac{BUT PDB} dataset. Best metrics are highlighted in bold type.}
\label{tab:brno}
\begin{tabular}{|c|ccccc|}
\hline
Method & \begin{tabular}[c]{@{}c@{}}Accuracy \\ (\%)\end{tabular} & \begin{tabular}[c]{@{}c@{}}Sensitivity\\  (\%)\end{tabular} & \begin{tabular}[c]{@{}c@{}}Specificity\\ (\%)\end{tabular} & \begin{tabular}[c]{@{}c@{}}Error\\ (Samples)\end{tabular} & \begin{tabular}[c]{@{}c@{}}Inference\\ Time (s)\end{tabular} \\ \hline
Peak   & 79.8 & 94.8 & 80.9 & \textbf{4.3} & 57.6 \\
CWT    & 76.8 & 83.8 & 84.3 & 30.5 & 120.7 \\ 
DWT    & 75.4 & \textbf{99.0} & 75.1 & 7.9 & 50.1 \\\hline
KEED   & \textbf{83.9} & 89.7 & \textbf{87.7} & 4.9 & \textbf{0.6} \\ \hline
\end{tabular}
\end{table}

While \ac{KEED} outperforms the performance of the WT-based algorithm, the overall effectiveness decreases compared to the previous experiment. It is worth noting that KEED’s optimization was based on a relatively small number of instances (2706). The presence of certain pathology in the evaluation dataset that have not been seen by the model during the optimization process may be causing the model to fail in its task. To address this, introducing a sufficient number of annotated instances into the training dataset is expected to resolve the issue. In contrast, adjusting various parameters within the WT algorithm would be necessary.

\subsection{Flexibility of \ac{KEED}'s proposed framework.}
\ac{KEED}, in a natural manner, allows users to adjust the probability threshold for classifying keypoints as either present or absent. While the proposed value for $\lambda$ is 0.4 for P wave detection due to its superior overall accuracy, it is essential to recognize that distinct clinical contexts may prioritize either detecting the presence or absence of P-Waves differently.

\begin{figure}[H]
\begin{center}
\centerline{\fbox{\includegraphics[width=\columnwidth]{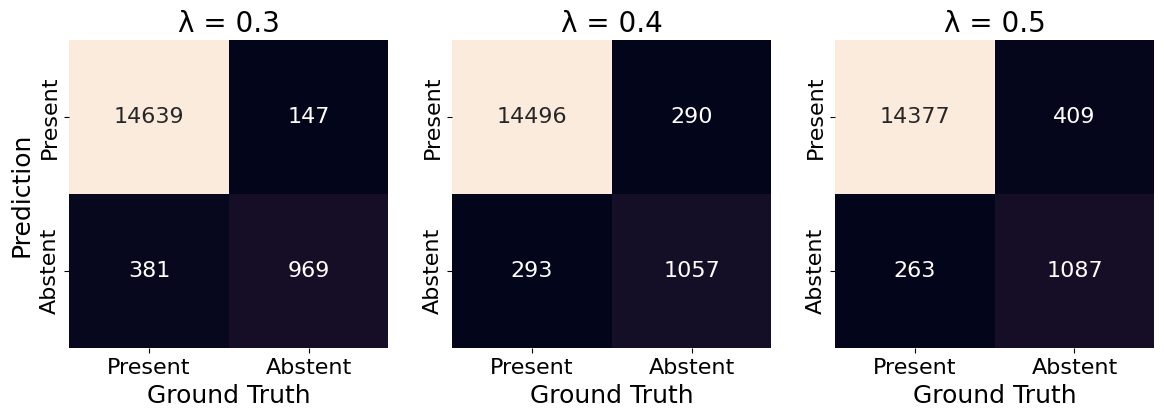}}}
\caption{Influence of $\lambda$ value in False Positives/False Negatives Trade-off.}
\label{fig:cms}
\end{center}
\end{figure}

Figure~\ref{fig:cms} represents distinct confusion matrices regarding distinct $\lambda$ values. It can be seen that increasing this $\lambda$ value leads to reducing the amount of False Negatives, i.e., whenever the model predicts the presence of a P-wave that was absent within the R-R interval. However, the number of False Positives, i.e., whenever the model does not identify the P-wave that was present within the R-R interval, increases accordingly.

\subsection{Discussion of the results}
In Table~\ref{tab:mit-afib-p}, the results highlight \ac{KEED}’s efficiency, surpassing existing \ac{SOTA} methods significantly and achieving substantial reductions in inference time—ranging from 52x to 703x faster. Meanwhile, Table~\ref{tab:brno} demonstrates that although the proposed method still outperforms other metrics, its performance slightly declines compared to the previous experiment. Considering the limited number of annotated instances used during training and the inclusion of previously unseen cardiac arrhythmia recordings, it is reasonable to expect that this issue will be resolved as more annotated data becomes available. Additionally, Figure~\ref{fig:cms} illustrates the trade-offs between specificity and sensitivity for different $\lambda$ values, showcasing the method’s flexibility to specific clinical requirements.

\section{Conclusions}
This study presents compelling evidence elucidating the tendency of addressing the signal delineation task form a sample-to-sample strategy. Instead, we introduce \ac{KEED} as a novel \ac{DL} method, which stands apart from these conventional methods by addressing the signal delineation challenge using a keypoints estimation approach. By doing this, we align the outcome of the model with the expected by the clinicians but also achieve significantly better results compared with \ac{SOTA} signal processing methods decreasing drastically the inference time. In addition, \ac{KEED} offers a flexible framework in which the false positives vs false negatives trade-off can be optimized depending the particular clinical specifications by just fine-tuning the the $\lambda$ value. 

\section{Limitations}
Only P-wave identification has been evaluated when assessing \ac{KEED}'s performance, since the lack of availability of alternative datasets to the one used during training with T-wave annotations. However, a comparable level of performance can be expected between  both P and T wave due to no particular considerations are incorporated to the model for identification of these two waves. In addition, we believe we have mitigated this absence of experiment by involving multiple datasets in the P-wave evaluation.

\bibliographystyle{plain} 
\bibliography{bib} 
\end{document}

%% file: acro.tex
\usepackage{acronym}
\usepackage[acronym]{glossaries}

\newacro{KEED}{Keypoint Estimation for Electrocardiogram Delineation} 
\newacro{ML}{Machine Learning}
\newacro{ECG}{electrocardiogram} 
\newacro{AFib}{Atrial Fibrillation}
\newacro{WT}{Wavelet Transform}
\newacro{DL}{Deep Learning}
\newacro{QT}{QT Database}
\newacro{MIT-PWave}{MIT-BIH Arrhythmia Database P-Wave Annotations}
\newacro{BUT PDB}{Brno University of Technology ECG Signal Database with Annotations of P Wave}
\newacro{SOTA}{state-of-the-art}